\documentclass[aip,pop,groupedaddress,floatfix,preprint]{revtex4-1}
\usepackage{microtype}
\usepackage{physics}
\usepackage{amsmath}
\usepackage{graphicx}
\usepackage{subfigure}
\usepackage{hyperref}
\usepackage{booktabs}
\usepackage{bm}

\DeclareMathOperator{\erfc}{erfc}

\begin{document}
\title{Simulations of self-accelerating electron phase space holes in an applied electric field} 
\author{Ran Guo}
\thanks{Author to whom correspondence should be addressed}
\email{rguo@cauc.edu.cn}
\affiliation{Department of Physics, College of Science, Civil Aviation University of China, Tianjin 300300, China}
\begin{abstract}
    The self-acceleration of electron phase space holes in an applied electric field is investigated via one-dimensional electrostatic Vlasov simulations.
    The electron holes (EHs) are initialized in a self-consistent manner with immobile ions, and the ion response is enabled at the beginning of simulations.
    A benchmark simulation is conducted to confirm the EH self-acceleration in the absence of the external electric field.
    Then, we investigate the EH behaviors by applying the uniform and sinusoidal electric fields, respectively.
    The effects of different strengths and durations of these external electric fields are studied.
    It is found that the uniform electric field applied in the direction of the self-acceleration can delay the onset of this process and change the final speed of EHs.
    The applied sinusoidal electric field can fix the EHs at their initial positions and suppress the self-acceleration if the electric field amplitude and duration are appropriate.
    In addition, it is observed that these external electric fields can induce the splitting of EHs and the generation of secondary EHs. 
    The physical mechanisms of these phenomena are discussed in detail.
\end{abstract}
\maketitle

\section{Introduction}
\label{sec:intro}
Electron phase space holes are one of the possible theoretical models for the solitary structures commonly observed in space plasmas 
\cite{Norgren2015, Graham2016, Mozer2018, Vasko2020a, Kamaletdinov2021, Norgren2022, Shaikh2025}
and laboratory plasmas. \cite{Sarri2010, Lefebvre2010}
The fast solitary waves observed in the auroral zone were inferred to be electron holes (EHs) due to their large amplitudes, speeds exceeding the ion-acoustic mode, and electromagnetic signatures. \cite{Ergun1998}
Norgren \textit{et al.} \cite{Norgren2015} identified the slow solitary waves observed in the plasma sheet boundary layer as EHs and made the unambiguous statistical estimates of EH velocities and length scales. 
The observations at Earth's magnetopause \cite{Graham2016} showed that the EH speeds span a wide range from almost zero to close to the electron thermal velocity in the ion frame, suggesting that the EHs may be generated by various instabilities.
A depletion in electron phase space, a critical feature of EHs, was directly observed by the Magnetospheric Multiscale Satellite, and the trapped electron distribution was measured. \cite{Mozer2018}
In the theoretical studies, various self-consistent models of EHs were constructed, and their characteristics were extensively investigated. \cite{Bernstein1957, Schamel1971, Turikov1984, Goldman2007, Schamel2025}

The EH is a localized positive potential structure caused by the deficit of electrons in phase space, so it exhibits distinct characteristics compared to the ion-acoustic solitons. \cite{Hutchinson2017,Hutchinson2024}
Ghizzo \textit{et al.} \cite{Ghizzo1988} proved that the periodic EHs in two-stream plasmas are unstable and would coalesce into a single EH. 
Saeki and Genma \cite{Saeki1998} studied how the EHs, which are self-consistent and stable in the sense of immobile ions, interact with mobile ions.
Their simulations revealed that the ion motion could result in the EH disruption and the formation of coupled hole-solitons (CHS). 
Eliasson and Shukla \cite{Eliasson2004} reported that the EHs, initially stationary relative to the Maxwellian ions, would undergo self-acceleration to high speeds.
The explanation for this phenomenon is that the imbalanced ion reflections from the potential and ion jetting exert a net force on the solitary potential when the potential is disturbed by a small shift. \cite{Hutchinson2021a}
The hole kinematics theory can quantitatively describe the self-acceleration process and predict the final EH speeds, which are in good agreement with the simulations. \cite{Hutchinson2016,Zhou2016} 
The self-acceleration was also confirmed by the observations in the Magnetospheric Multiscale mission. \cite{Dong2023}

Due to the inherent self-acceleration, the EH seems unsuitable as a theoretical model for the slow solitary waves observed in space plasmas \cite{Khotyaintsev2010, Graham2016, Lotekar2020}, which propagate close to or below the ion-acoustic speed in the ion frame.
However, Hutchinson demonstrated that a slow EH can be stable and does not undergo self-acceleration 
if the background ions follow a double-humped distribution and the EH speed lies in the local minimum of the ion distribution. \cite{Hutchinson2021a}
This theory is supported by the observations. \cite{Kamaletdinov2021}
In addition, the slow solitary waves can also be explained by the soliton model. \cite{Lakhina2024, Guo2025}

As electric and magnetic fields are ubiquitous in space plasmas, \cite{Foukal1991,Marklund2009} the EH behaviors in applied fields are of interest.
Previous studies focused on the evolution of EHs in inhomogeneous magnetic fields. \cite{Vasko2016, Kuzichev2017}
Vasko \textit{et al.} \cite{Vasko2016} found that the trapped electrons are heated when EHs propagate into the stronger magnetic field region,
suggesting a possible generation mechanism for suprathermal electron fluxes.
Kuzichev \textit{et al.} \cite{Kuzichev2017} reported that the EH speeds are modified when the EHs travel into the magnetic fields of different strengths, and the acceleration rates depend on the field gradients.
Moreover, the external electric field can also have significant effects on plasmas, such as causing anomalous transport, \cite{Hamberger1972} driving instabilities, \cite{Reiter1967, Beving2023} heating the plasmas parallel and perpendicular to the magnetic field, \cite{Vranjes2009} and changing the shape of distribution functions. \cite{Swanson2003, Anderegg2009, Zanelli2025}
It is worth noting that the EH evolution in inhomogeneous plasmas was investigated by modeling the density gradient via an external forcing term in the Vlasov equation, \cite{Vasko2017} which is mathematically analogous to the treatment of the external electric field in the present study.
Their results showed that EHs can be accelerated (decelerated) and narrowed (widened) when propagating through inhomogeneous plasmas.
Therefore, it is reasonable to expect that the self-acceleration of EHs may be greatly affected by the applied electric field.
To enable the self-acceleration, the ion response is included in the present study, and the non-stationary electric field is also considered, differing from the simulation setup in the previous study. \cite{Vasko2017}

In this work, we conduct one-dimensional electrostatic Vlasov simulations to study the self-acceleration processes of EHs when the external electric fields are applied.
In Sec. \ref{sec:sim}, the simulation model is described in detail, and the parameter setup is provided.
The simulation results are presented in Sec. \ref{sec:results}.
We perform a benchmark simulation of EH self-acceleration in the absence of background fields to validate the codes in Sec. \ref{sec:sa}.
Then, the different behaviors of EHs are investigated by applying the uniform and sinusoidal electric fields in Secs. \ref{sec:uniform-e} and \ref{sec:sin-e}, respectively.
Finally, a summary is given in Sec. \ref{sec:summary}.

\section{Simulation Model}
\label{sec:sim}

We consider a one-dimensional electrostatic plasma in the frame moving with the solitary EH.
For convenience, the dimensionless variables are employed in this study.
The length is normalized by the electron Debye length $\lambda_{De} = \sqrt{\epsilon_0 k_B T_e/(n_0 e^2)}$, 
the velocity by the electron thermal velocity $v_{te} = \sqrt{k_B T_e/m_e}$, 
the number density by $n_0$ which is the undisturbed density at $x\rightarrow \pm \infty$ without the applied electric field,
the potential by $k_B T_e/e$, 
and the time by the inverse of the electron plasma frequency $\omega_{pe}^{-1} = \sqrt{m_e \epsilon_0/(n_0 e^2)}$. 

The initial EH is constructed by a self-consistent scheme in the case of immobile ions.
The solitary potential is assumed to be,
\begin{equation}
    \phi(x, t=0) = \psi \sech^2\left(\frac{x-L/2}{\Delta}\right),
    \label{eq:phi}
\end{equation}
where $\psi$ is the potential amplitude, $\Delta$ characterizes the EH width, and $L$ is the simulation box length.
The electrons passing the solitary potential are presumed to follow the non-drifting Maxwellian distribution,
\begin{equation}
    f_{e,p}(w, t=0) = \frac{1}{\sqrt{2\pi}} e^{-w}, \quad \text{for } w > 0,
    \label{eq:fe0-p}
\end{equation}
where $w = v^2/2 - \phi$ is the total energy of the electron.
The trapped electron distribution is set to be self-consistent with the solitary potential \eqref{eq:phi}, the passing electron distribution \eqref{eq:fe0-p}, and the background uniform and immobile ions.
Its analytical form is given by Turikov via the BGK integral method, \cite{Turikov1984}
\begin{equation}
    f_{e,t}(w, t=0) = \frac{1}{\sqrt{2 \pi}} e^{-w}\erfc(\sqrt{-w}) + \frac{8\sqrt{-w}}{\sqrt{2} \pi \Delta^2}\left(\frac{2w}{\psi}+1\right), \quad \text{for } w < 0,
    \label{eq:fe0-t}
\end{equation}
where $\erfc(z) = (2/\sqrt{\pi}) \int_z^\infty \exp(-t^2) \dd{t}$ is the complementary error function. \cite{Olver2010}
In addition, the ions are initially Maxwellian and spatially uniform,
\begin{equation}
    f_i(x, v, t=0) = \frac{1}{\sqrt{2\pi} v_{ti}} \exp\left[-\frac{(v-u_i)^2}{2 v_{ti}^2}\right],
    \label{eq:fi0}
\end{equation}
with the ion thermal velocity expressed in dimensionless form as $v_{ti} = \sqrt{k_B T_i/m_i}/ v_{te}$ (i.e., in units of $v_{te}$).

\begin{figure}
    \centering
    \includegraphics[width=0.5\textwidth]{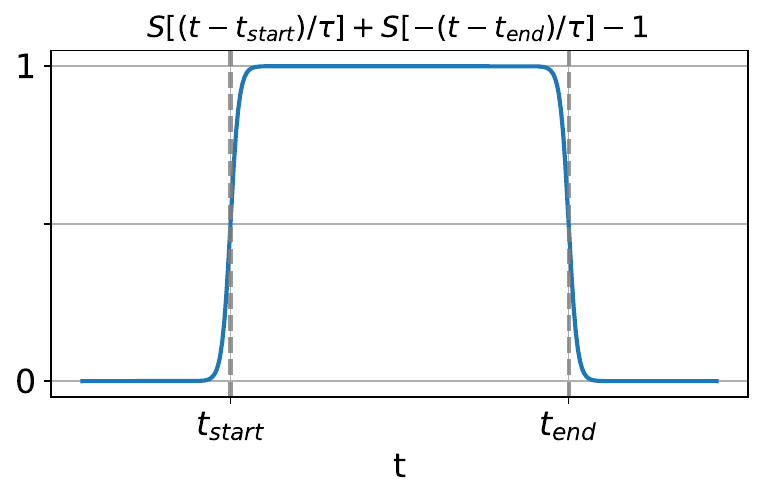}
    \caption{Schematic of the term in the square bracket of Eq. \eqref{eq:Ea-t}.}
    \label{fig:sigmoid}
\end{figure}

When an external electric field is imposed and the ion response is turned on, the plasma evolution is governed by the dimensionless Vlasov-Poisson equations,
\begin{equation}
    \pdv{f_e}{t} + v\pdv{f_e}{x} + \pdv{(\phi_0+\phi_{a})}{x}\pdv{f_e}{v} = 0,
    \label{eq:vlasov-e}
\end{equation}
\begin{equation}
    \pdv{f_i}{t} + v\pdv{f_i}{x} - \frac{1}{\mu}\pdv{(\phi_0+\phi_{a})}{x}\pdv{f_i}{v} = 0,
    \label{eq:vlasov-i}
\end{equation}
\begin{equation}
    \pdv[2]{\phi_0}{x} = \int f_e \dd v - \int f_i \dd v,
    \label{eq:poisson}
\end{equation}
where $\mu=m_i/m_e$ is the ion-to-electron mass ratio.
The total potential $\phi$ is decomposed into the potential $\phi_0$ generated by the locally non-neutral charge and the applied potential $\phi_a$.
The latter is associated with the applied electric field $E_a = -\pdv*{\phi_a}{x}$.
In the simulation, the external field is added to or removed from the plasma adiabatically to avoid unphysical perturbations,
\begin{equation}
    E_a(x,t) = \hat{E}_{a}(x) \left[S\left(\frac{t-t_{start}}{\tau}\right) + S\left(-\frac{t-t_{end}}{\tau}\right) - 1 \right],
    \label{eq:Ea-t}
\end{equation}
where $\hat{E}_a$ is the target profile of the applied field, $S(z) = 1/(1+e^{-z})$ is the sigmoid function,
$t_{start}$ ($t_{end}$) is the time to add (remove) the external electric field, and $\tau = 0.1$ is the ramping time.
The term in the square bracket on the right side of Eq. \eqref{eq:Ea-t} is shown in Fig. \ref{fig:sigmoid}.

In this work, the simulation codes solve the Vlasov equations via the time-splitting algorithm with cubic spline interpolation \cite{Cheng1976} and the Poisson equation through the tridiagonal matrix algorithm. \cite{Press2007}
These codes have been validated by simulating Landau damping and two-stream ion-ion instability. \cite{Guo2021a, Guo2025}
In the current simulations, the position domain $[0, L]$ with $L = 60 \pi \lambda_{De}$ is discretized with $N_x = 4000$ grid points, and the periodic boundary conditions are employed.
The electron velocity space $[-10v_{te},10v_{te}]$ and the ion velocity space $[-10v_{ti}, 10v_{ti}]$ are both resolved with $N_{v} = 2000$ grid points.
The velocity distributions are truncated to zero when the velocity is outside the velocity domain.
The time step is $\dd t=0.05\omega_{pe}^{-1}$, and the maximum simulation time is $1500\omega_{pe}^{-1}$.
The ion-to-electron mass ratio is chosen as $\mu=1836$, the ion temperature is $T_i = 5 T_e$, and the ion drift speed is $u_i = -0.1v_{te}$.
The initial amplitude and width of the solitary potential \eqref{eq:phi} are set to $\psi = 0.1k_B T_e/e$ and $\Delta = 5.0\lambda_{De}$, respectively, in accordance with the observed parameter ranges. \cite{Lotekar2020}

\section{Results}
\label{sec:results}
\subsection{Benchmark: Self-acceleration of EHs}
\label{sec:sa}
The benchmark simulation is performed to confirm the EH self-acceleration in the absence of the applied electric field, i.e., by setting 
\begin{equation}
    \hat{E}_a(x) = 0 
    \label{eq:sa-benchmark-Ea}
\end{equation}
in Eq. \eqref{eq:Ea-t}.

\begin{figure}
    \centering
    \includegraphics[width=0.5\textwidth]{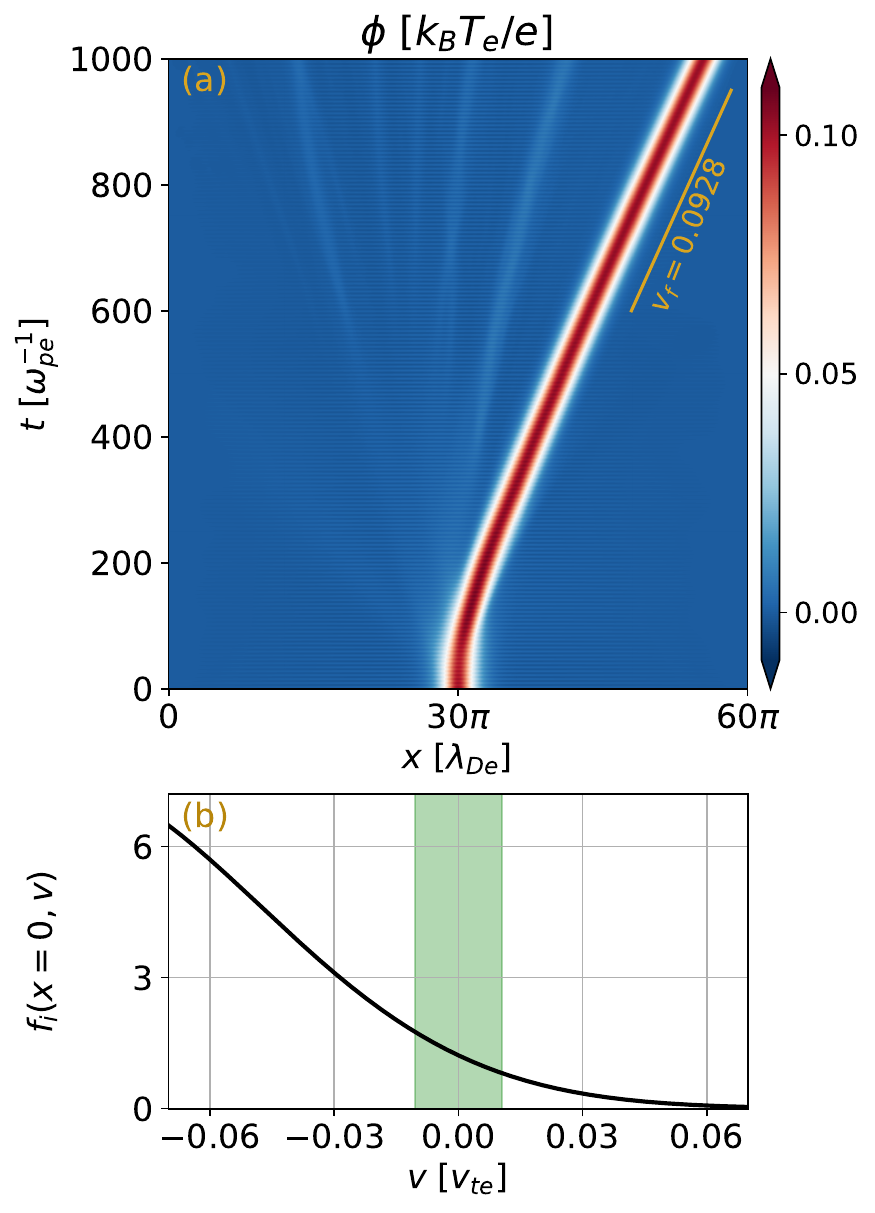}
    \caption{
        (a) The potential evolution in the benchmark simulation of self-acceleration, where the applied electric field is absent.
        The final EH speed $v_{f}$ is measured by the linear fitting of the EH position from $t=600\omega_{pe}^{-1}$ to $t=950\omega_{pe}^{-1}$.
        (b) The initial velocity distribution of ions.
        The green region denotes the velocity range of the reflected ions, determined by the initial amplitude of the solitary potential $\psi = 0.1 k_B T_e/e$. 
    }
    \label{fig:sa-benchmark}
\end{figure}

The potential evolution is illustrated in Fig. \ref{fig:sa-benchmark}(a).
At the initial stage, because of the negative bulk speed of ions, the ion distribution \eqref{eq:fi0} has a negative slope at the EH speed, 
representing that more ions move toward the negative $x$-direction as presented in Fig. \ref{fig:sa-benchmark}(b).
Therefore, the ion reflections occur more frequently on the right side of the solitary potential.
These reflected ions equivalently experience a net force toward the right, while the EH is subjected to a reaction force in the opposite direction.
As its effective mass is negative,\cite{Hutchinson2021a} the EH is accelerated to the right in this simulation.
Figure \ref{fig:sa-benchmark}(a) depicts that the self-acceleration occurs rapidly after the initialization,
and the EH is accelerated to the final speed $v_{f} = 0.0928 v_{te}$ after roughly $t=200\omega_{pe}^{-1}$. 
Strictly speaking, the final state of the solitary potential should be a CHS due to the positive perturbation of ion density, but we still refer to it as an EH for convenience. 

\subsection{Uniform electric field}
\label{sec:uniform-e}
\begin{figure}
    \centering
    \includegraphics[width=0.5\textwidth]{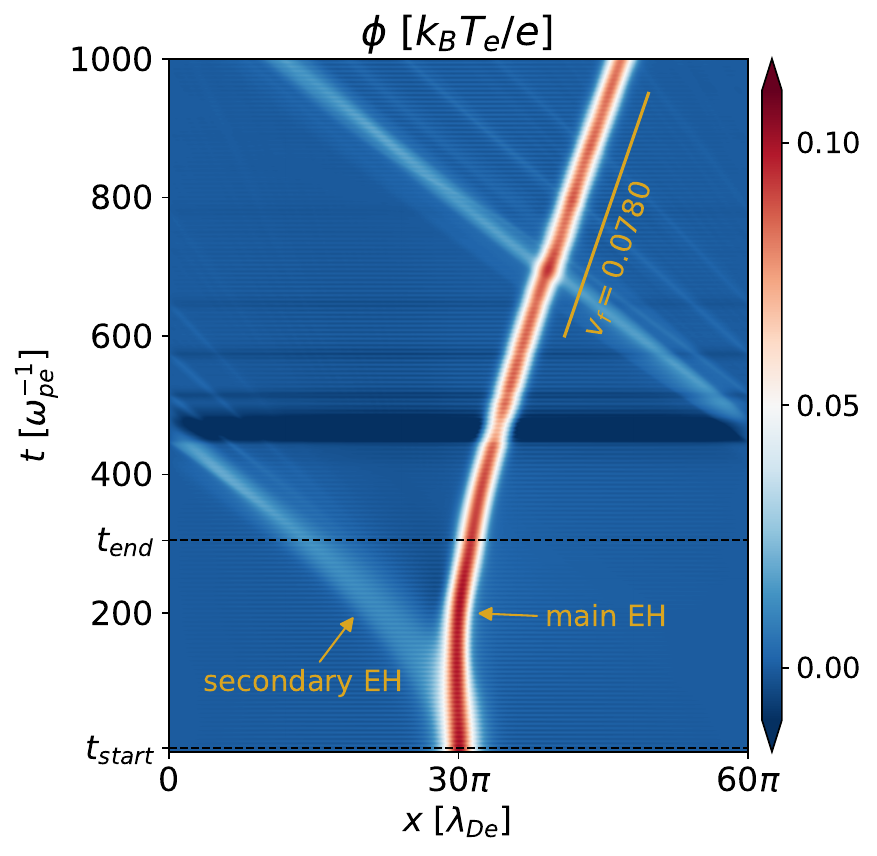}
    \caption{
        The potential evolution for the case of applying the uniform electric field with $\hat{E}_a = 0.0008k_BT_e/(e\lambda_{De})$ from $t_{start}=5\omega_{pe}^{-1}$ to $t_{end}=305\omega_{pe}^{-1}$.
        The start and end times of the applied field are marked by the horizontal dashed lines.
        The final EH speed $v_{f}$ is determined by the linear fitting of the EH position from $t=600\omega_{pe}^{-1}$ to $t=950\omega_{pe}^{-1}$.
    }
    \label{fig:sa-uniform-E00008}
\end{figure}
In this subsection, we apply a uniform electric field directed to the right,
i.e., by setting 
\begin{equation}
    \hat{E}_a(x) = C 
    \label{eq:uniform-Ea}
\end{equation}
in Eq. \eqref{eq:Ea-t}, where $C$ is a positive constant. 

Figure \ref{fig:sa-uniform-E00008} exhibits the potential evolution if the uniform field $\hat{E}_a = 0.0008k_BT_e/(e\lambda_{De})$ is imposed during $t=5\omega_{pe}^{-1}$ to $t=305\omega_{pe}^{-1}$.
It shows that the onset of the self-acceleration is delayed and the final EH speed is reduced in comparison with Fig. \ref{fig:sa-benchmark}(a).
This behavior can be simply explained as follows.
From the benchmark simulation, we can conclude that the self-acceleration is caused by the leftward force exerted on the EH that possesses a negative effective mass.
On the one hand, the background field $\hat{E}_a > 0$ exerts a rightward force on the EH, thus resisting the self-acceleration process.
On the other hand, because the uniform electric field exerts the same force on all ions, it shifts the velocity distribution of ions without modifying its shape.
Therefore, the imbalanced ion reflections still exist after the electric field is removed.
As a result, the EH still undergoes the self-acceleration but finally reaches a lower speed.

\begin{figure}
    \centering
    \includegraphics[width=0.5\textwidth]{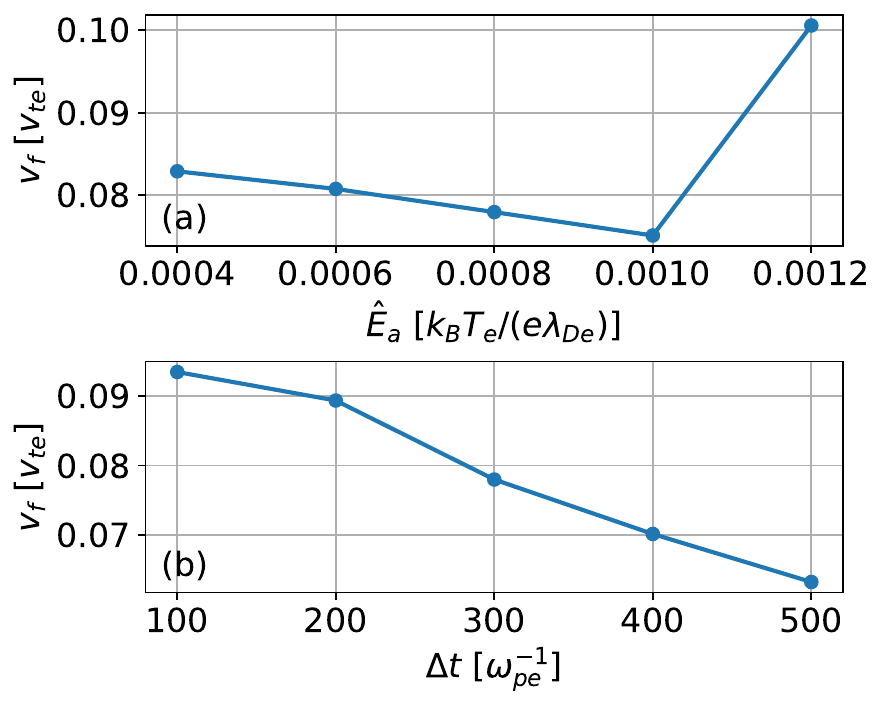}
    \caption{
        The final EH speeds $v_{f}$ by varying (a) the magnitude $\hat{E}_a$ and (b) the duration $\Delta t = t_{end} - t_{start}$ of the applied uniform electric field.
        The electric field is added at the fixed time $t_{start} = 5\omega_{pe}^{-1}$.
        In the upper panel, $t_{end}$ is set to $305\omega_{pe}^{-1}$, and $\hat{E}_a$ is varied from $0.0004k_BT_e/(e\lambda_{De})$ to $0.0012k_BT_e/(e\lambda_{De})$. 
        In the lower panel, $\hat{E}_a$ is fixed at $0.0008 k_BT_e/(e\lambda_{De})$, and $t_{end}$ is varied from $105\omega_{pe}^{-1}$ to $505\omega_{pe}^{-1}$.
        All the final speeds are calculated by the linear fitting of the EH position from $t=600\omega_{pe}^{-1}$ to $t=950\omega_{pe}^{-1}$.
    }
    \label{fig:vfinal-uniform-E}
\end{figure}
\begin{figure}
    \centering
    \includegraphics[width=0.5\textwidth]{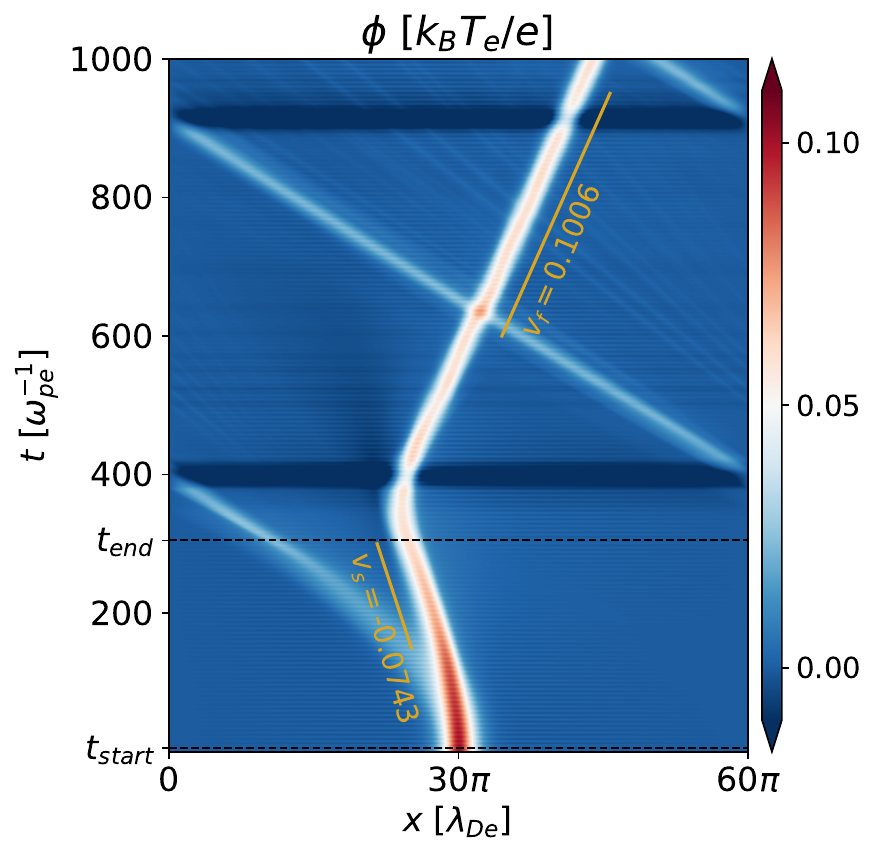}
    \caption{
        The potential evolution for the case of applying the uniform electric field with $\hat{E}_a = 0.0012k_BT_e/(e\lambda_{De})$ from $t_{start}=5\omega_{pe}^{-1}$ to $t_{end}=305\omega_{pe}^{-1}$.
        The start and end times of the applied electric field are denoted by the horizontal dashed lines.
        The EH speed $v_{s}$ is fitted from $t=150\omega_{pe}^{-1}$ to $t=300\omega_{pe}^{-1}$,
        while the final speed $v_{f}$ from $t=600\omega_{pe}^{-1}$ to $t=950\omega_{pe}^{-1}$.
    }
    \label{fig:sa-uniform-E00012}
\end{figure}

Figure \ref{fig:vfinal-uniform-E} illustrates the effects of different strengths and durations of the applied uniform electric field on the final EH speed.
In the upper panel, one finds that the final EH speed decreases with the increasing field intensity $\hat{E}_a$ from $0.0004k_BT_e/(e\lambda_{De})$ to $0.0010k_BT_e/(e\lambda_{De})$.
However, the final EH speed increases when the electric field is further enhanced to $0.0012k_BT_e/(e\lambda_{De})$. 
The potential evolution of this case is depicted in Fig. \ref{fig:sa-uniform-E00012}.
It is evident that the background electric field is so strong that the EH is leftward accelerated to $v_s$ before the removal of the electric field, where $v_s = -0.0743 v_{te}$ is fitted from the simulation.
This is different from the other cases, e.g., the one shown in Fig. \ref{fig:sa-uniform-E00008}, where the measured EH speed is $0.0398 v_{te}$ before the external electric field is removed.
It is worth noting that the absolute value of the slope $|\partial_v f_i|_{v=-0.0743 v_{te}} = 63.9$ is larger than $|\partial_v f_i|_{v=0.0398 v_{te}} = 10.9$.
In other words, compared to the case in Fig. \ref{fig:sa-uniform-E00008}, the EH in the present case is located in a region of the ion velocity distribution where the slope is steeper. 
Guillevic \textit{et al.} reported that a steeper slope of the ion distribution at the EH speed leads to a stronger self-acceleration, \cite{Guillevic2025} which could be attributed to the greater imbalance in ion reflections represented by the steeper slope $\partial_v f_i$ at the EH speed.
Consequently, after the electric field is removed, the EH is accelerated to a higher final speed. 
Besides, the lower panel of Fig. \ref{fig:vfinal-uniform-E} shows that the final EH speed diminishes with the increasing duration.

\begin{figure}
    \centering
    \includegraphics[width=0.5\textwidth]{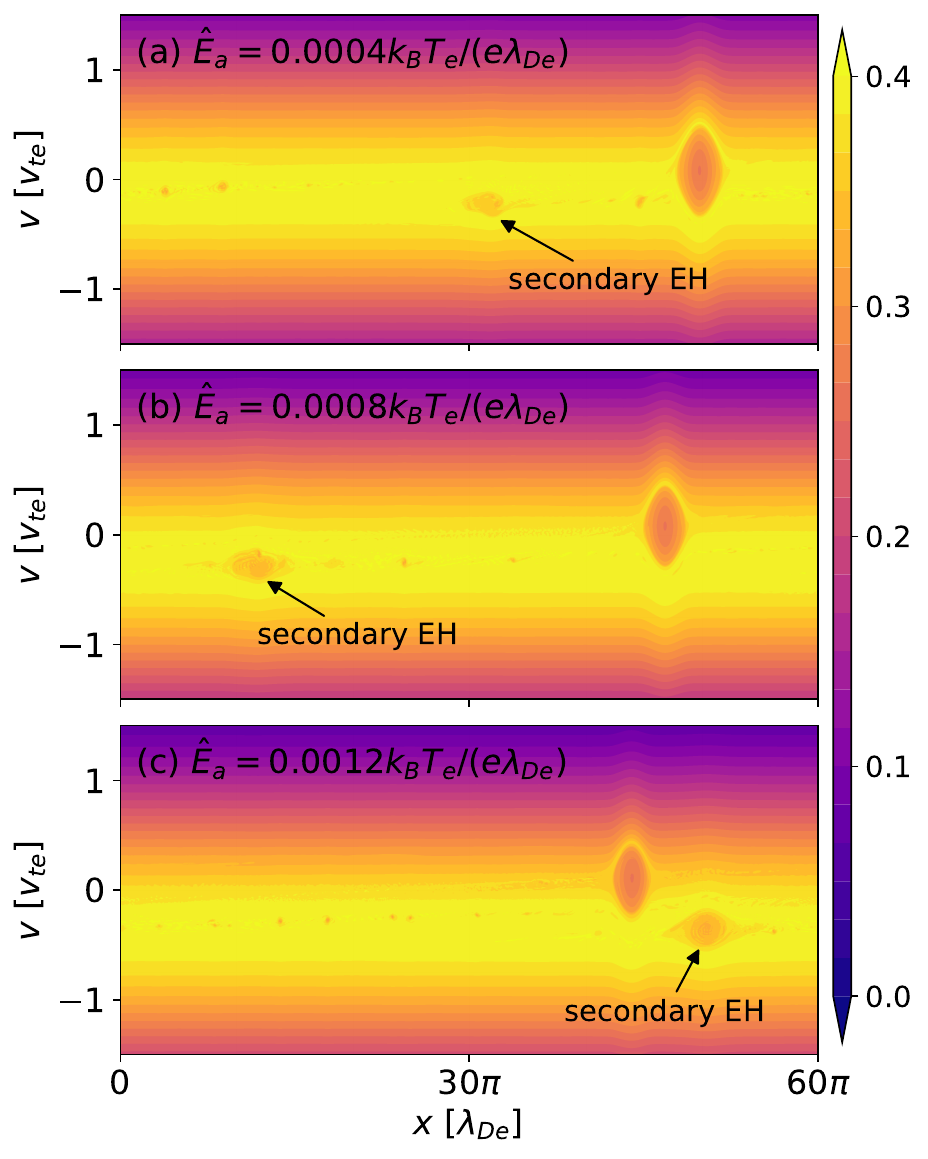}
    \caption{
        The electron phase space at $t=1000\omega_{pe}^{-1}$ for the applied uniform electric fields (a) $\hat{E}_a = 0.0004k_BT_e/(e\lambda_{De})$, (b) $\hat{E}_a = 0.0008k_BT_e/(e\lambda_{De})$, and (c) $\hat{E}_a = 0.0012k_BT_e/(e\lambda_{De})$.
    }
    \label{fig:sa-uniform-E-com-fe}
\end{figure}

In addition to weakening the self-acceleration, the applied uniform electric field also induces the splitting of secondary EHs from the main EH, as marked in Fig. \ref{fig:sa-uniform-E00008}.
\begin{figure}
    \centering
    \includegraphics[width=0.5\textwidth]{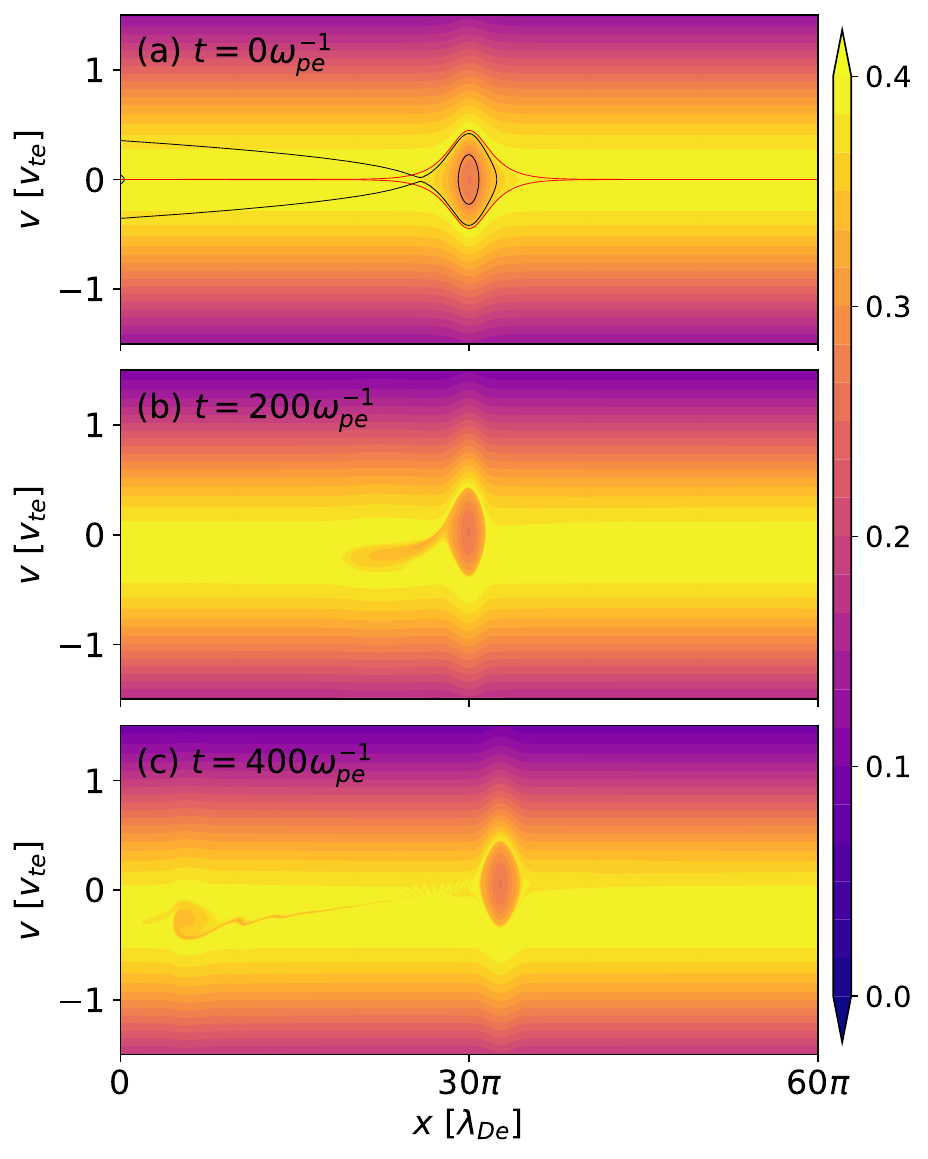}
    \caption{
        The time evolution of the electron phase space with the applied uniform electric field $\hat{E}_a = 0.0008k_BT_e/(e\lambda_{De})$ at (a) $t=0\omega_{pe}^{-1}$, (b) $t=200\omega_{pe}^{-1}$, and (c) $t=400\omega_{pe}^{-1}$.
        The red solid line denotes the separatrix between the trapped and passing electrons without the applied electric field.
        The black solid line represents the contours of constant energy after the application of the uniform electric field.
    }
    \label{fig:sa-uniform-E00008-fe-evo}
\end{figure}
The detailed separation process is illustrated by the time evolution of the electron phase space in Fig. \ref{fig:sa-uniform-E00008-fe-evo}.
Panel (a) shows the initial electron phase space, where the red solid line denotes the separatrix between the trapped and passing electrons in the absence of an applied electric field.
The application of the uniform electric field modifies the electron energy to $w = v^2/2 - \phi_0 + \hat{E}_a x$, and the corresponding contours of constant energy immediately after the field application are represented by the black solid lines. 
Although these initial energy contours evolve over time, they can be used as a qualitative reference for understanding the system's behavior.
As electrons move approximately along these initial constant-energy contours, the originally trapped electrons close to the separatrix can escape from the EH under the influence of the applied electric field, while those in the center of the EH remain trapped.
Hence, a secondary EH is dragged out from the main EH and eventually moves in the direction opposite to that of the main EH, as shown in Figs. \ref{fig:sa-uniform-E00008-fe-evo}(b) and (c).
Additionally, the secondary EH increases in size when a stronger uniform electric field is applied, as shown in Fig. \ref{fig:sa-uniform-E-com-fe}. 

\subsection{Sinusoidal electric field}
\label{sec:sin-e}
The applied electric field considered here is set as a sinusoidal function,
\begin{equation}
    \hat{E}_a(x) = \mathcal{E}_a \sin(k x),
    \label{eq:sin-Ea}
\end{equation}
which corresponds to a periodic wave propagating with the same speed as the EH.
The wave number is fixed at $k\lambda_{De} = 1$, implying that there are $30$ wavelengths in the simulation box due to the setting $L = 60 \pi \lambda_{De}$.

\begin{figure}
    \centering
    \includegraphics[width=0.5\textwidth]{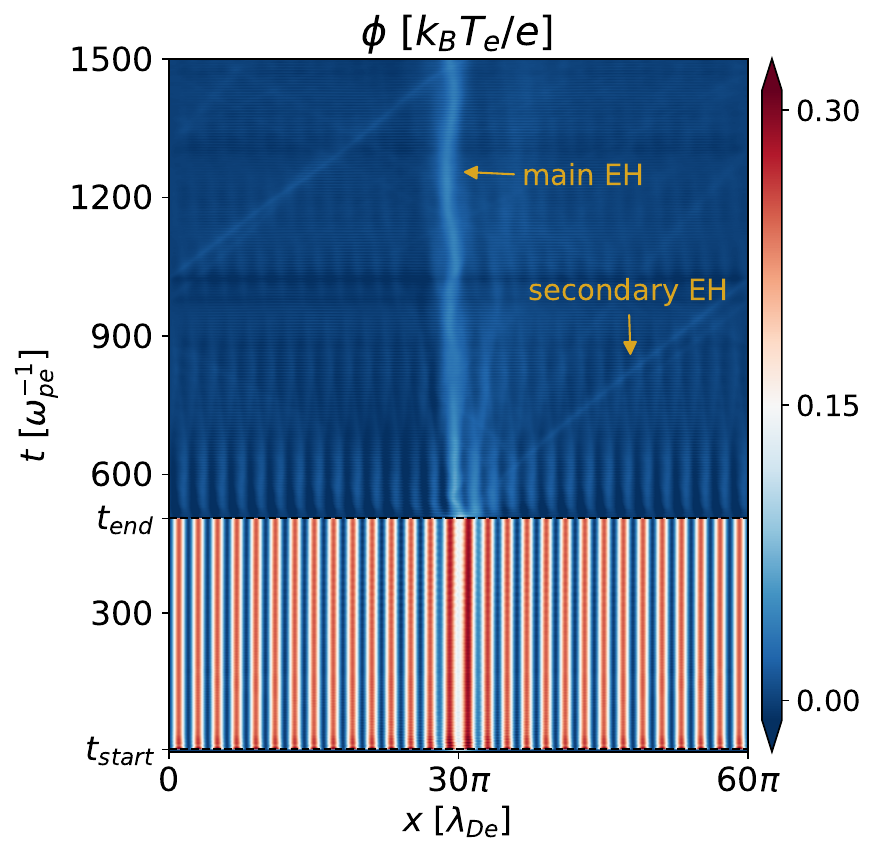}
    \caption{
        The potential evolution for the case of applying the sinusoidal electric field with $\mathcal{E}_a = 0.3 k_BT_e/(e\lambda_{De})$ from $t_{start}=5\omega_{pe}^{-1}$ to $t_{end}=505\omega_{pe}^{-1}$.
        The start and end times of the applied field are denoted by the horizontal dashed lines.
    }
    \label{fig:sa-sin-E03}
\end{figure}

\begin{figure*}
    \centering
    \includegraphics[width=\textwidth]{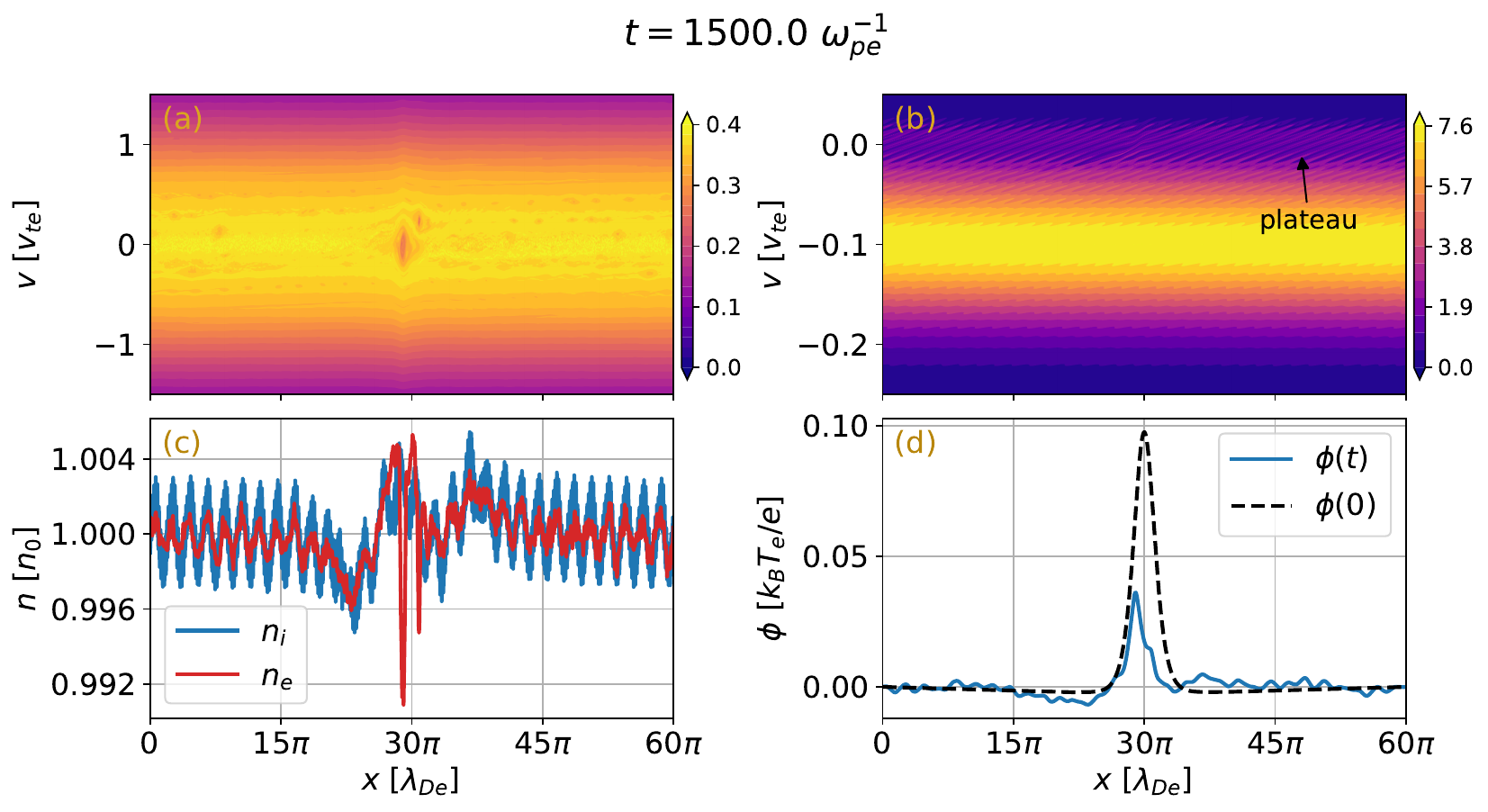}
    \caption{
        The final state of the EH at $t=1500\omega_{pe}^{-1}$ for the case of applying the sine electric field with $\mathcal{E}_a = 0.3 k_BT_e/(e\lambda_{De})$ from $t_{start}=5\omega_{pe}^{-1}$ to $t_{end}=505\omega_{pe}^{-1}$, including (a) the electron phase space, (b) the ion phase space, (c) the number density, and (d) the potential.
    }
    \label{fig:phase-space-final-sinE03}
\end{figure*}

In this setup, an interesting phenomenon is the stationary behavior of the main EH after the removal of the electric field. 
Figure \ref{fig:sa-sin-E03} plots the potential evolution for the case of $\mathcal{E}_a = 0.3k_BT_e/(e\lambda_{De})$ with $t_{start}=5\omega_{pe}^{-1}$ and $t_{end}=505\omega_{pe}^{-1}$.
Although the amplitude decreases, the main solitary potential does not undergo self-acceleration and remains near its initial position. 
As mentioned in Sec. \ref{sec:sa}, the self-acceleration is induced by the imbalanced ion reflections, which are described by the negative slope of the ion velocity distribution at the EH speed.
However, the sine electric field \eqref{eq:sin-Ea} is a wave propagating with the same speed as the EH, so it can flatten the ion velocity distribution at the EH speed and thus suppress the self-acceleration.
Figure \ref{fig:phase-space-final-sinE03} displays the electron and ion phase spaces, the number density, and the potential profile at $t=1500\omega_{pe}^{-1}$.
The ion phase space in Fig. \ref{fig:phase-space-final-sinE03}(b) exhibits a plateau around the EH speed.
\begin{figure}
    \centering
    \includegraphics[width=0.5\textwidth]{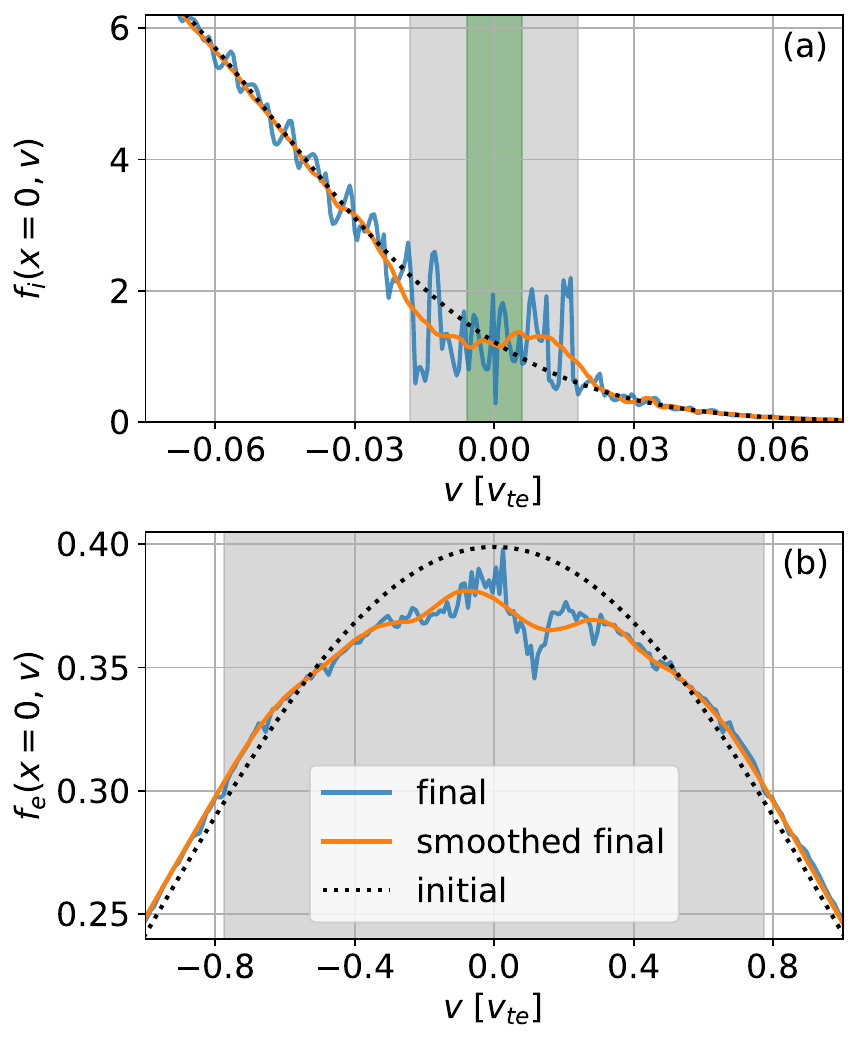}
    \caption{
        The (a) ion and (b) electron velocity distributions $f_{i,e}(x=0,v)$ for the case of applying the sine electric field with $\mathcal{E}_a = 0.3 k_BT_e/(e\lambda_{De})$ from $t_{start}=5\omega_{pe}^{-1}$ to $t_{end}=505\omega_{pe}^{-1}$.
        The blue lines mark the final velocity distributions at $t=1500\omega_{pe}^{-1}$, whereas the black dotted lines represent the initial distributions.
        The Savitzky-Golay filter with a window size of 80 and a polynomial degree of 7 is used to smooth the final velocity distributions, and the corresponding results are shown by the orange lines.
        The gray regions indicate the velocity ranges for ions and electrons, respectively, trapped in the external sine electric field.
        The green region in the upper panel indicates the velocity range for the reflected ions, which is determined by the potential amplitude $\psi_{fit} = 0.0323 k_B T_e/e$ fitted from the potential in Fig. \ref{fig:phase-space-final-sinE03}(d).
    }
    \label{fig:pdf-final-sinE03}
\end{figure}
More explicitly, the ion velocity distribution at $x=0$ is illustrated in Fig. \ref{fig:pdf-final-sinE03}(a),
where the blue line represents the final distribution at $t=1500\omega_{pe}^{-1}$ and the orange line is the smoothed result by using the Savitzky-Golay filter\cite{Press2007}. 
The green region denotes the velocity range for the reflected ions, which is determined by the fitted amplitude of the solitary potential $\psi_{fit} = 0.0323 k_B T_e/e$ from the final state plotted in Fig. \ref{fig:phase-space-final-sinE03}(d).
At the position $x=0$, the solitary potential vanishes, so the modification of the ion velocity distribution in Fig. \ref{fig:pdf-final-sinE03}(a) is solely attributed to the ion trapping in the applied electric field.
The gray region indicates the velocity range for ions trapped in the sinusoidal electric field, which is determined by its potential amplitude $\mathcal{E}_a/k = 0.3 k_B T_e/e$.
It is observed that the smoothed velocity distribution of ions features a plateau around $v=0$, and the reflected ions are distributed within this flat region.
Therefore, the ion reflections are balanced on both sides of the solitary potential, and the self-acceleration does not occur.
It should be emphasized that the formation of such an appropriate plateau in the final ion distribution serves as a necessary condition for suppressing the self-acceleration.
In other words, if a suitable plateau fails to form even in this final state, the self-acceleration cannot be effectively prevented.

In addition, the electron velocity distribution $f_e(x=0, v)$ is displayed in Fig. \ref{fig:pdf-final-sinE03}(b).
The final distribution deviates from the initial one due to the applied electric field, and the velocity range for electrons trapped in the background field is denoted by the gray region.
However, it does not exhibit an obvious plateau around the EH speed.
Thus, the ion distribution modified by the applied electric field is the main reason for the suppression of the EH self-acceleration.

\begin{figure*}
    \centering
    \includegraphics[width=\textwidth]{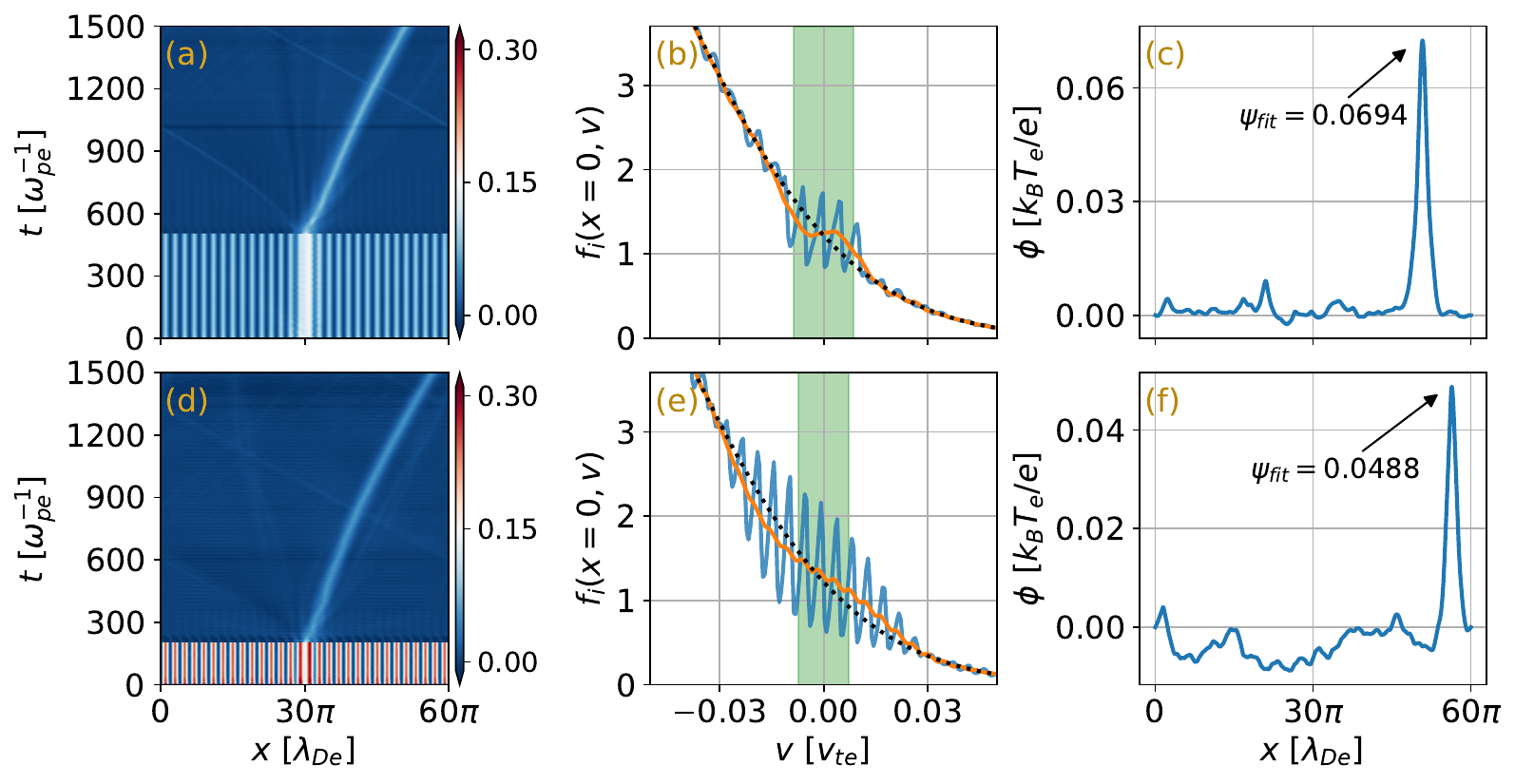}
    \caption{
        The comparisons by applying the sinusoidal electric field with $\mathcal{E}_a = 0.1 k_BT_e/(e\lambda_{De})$ from $t=5 \omega_{pe}^{-1}$ to $505\omega_{pe}^{-1}$ (panels (a)-(c)), and $\mathcal{E}_a = 0.3k_BT_e/(e\lambda_{De})$ from $t=5 \omega_{pe}^{-1}$ to $205\omega_{pe}^{-1}$ (panels (d)-(f)).
        From left to right, the columns depict the potential evolutions, the final velocity distributions of ions, and the final potential profiles.
        The green regions in (b) and (e) denote the velocity ranges for the reflected ions, which are derived by the fitted potential amplitudes in (c) and (f), respectively.
    }
    \label{fig:sinE-comparison}
\end{figure*}

According to the above analysis, the applied electric field should be strong enough and long-lived to flatten a large enough plateau in the ion velocity distribution, and consequently eliminate the self-acceleration.
Two comparative simulations are conducted by applying the sinusoidal electric field with $\mathcal{E}_a = 0.1 k_BT_e/(e\lambda_{De})$ from $t=0 \omega_{pe}^{-1}$ to $505\omega_{pe}^{-1}$ and $\mathcal{E}_a = 0.3k_BT_e/(e\lambda_{De})$ from $t=0 \omega_{pe}^{-1}$ to $205\omega_{pe}^{-1}$, respectively.
The results are plotted in Fig. \ref{fig:sinE-comparison}.
Figure \ref{fig:sinE-comparison}(b) illustrates that the plateau in the velocity distribution is not wide enough to cover the range of the reflected ions due to the weak field intensity $\mathcal{E}_a = 0.1 k_BT_e/(e\lambda_{De})$.
Moreover, the short-lived electric field from $t=5 \omega_{pe}^{-1}$ to $205\omega_{pe}^{-1}$ cannot form the plateau in the velocity distribution, as shown in Fig. \ref{fig:sinE-comparison}(e).
Therefore, both the above two cases undergo the self-acceleration, as illustrated in Figs. \ref{fig:sinE-comparison}(a) and (d).

The elimination of the EH self-acceleration by the external sinusoidal waves can be achieved across a certain range of EH speeds and sinusoidal wave phase speeds relative to the EH, 
as long as the applied electric field flattens the range containing the reflected ions in the ion velocity distribution.
For verification, we conduct the simulations by changing the ion drift speeds (i.e., the negative of the EH speeds relative to the ions) to $u_i = -0.05 v_{te}$, $-0.075 v_{te}$, $-0.125 v_{te}$, and $-0.15 v_{te}$, 
while keeping the other parameters identical to those in Fig. \ref{fig:sa-sin-E03} (in which $u_i = -0.1 v_{te}$).
None of these simulations exhibit significant self-acceleration.
Furthermore, the cases with the sinusoidal wave velocities $\omega/k = -0.04 v_{te}$, $-0.02 v_{te}$, $0.02 v_{te}$, and $0.04 v_{te}$ are also simulated (for reference, $\omega/k = 0$ in our previous simulations).
In these cases, the EH is trapped by the external sinusoidal wave and forced to propagate at the wave phase velocity.
After the external field is removed, no additional EH self-acceleration is observed, which confirms the above analysis.

\begin{figure}
    \centering
    \includegraphics[width=0.5\textwidth]{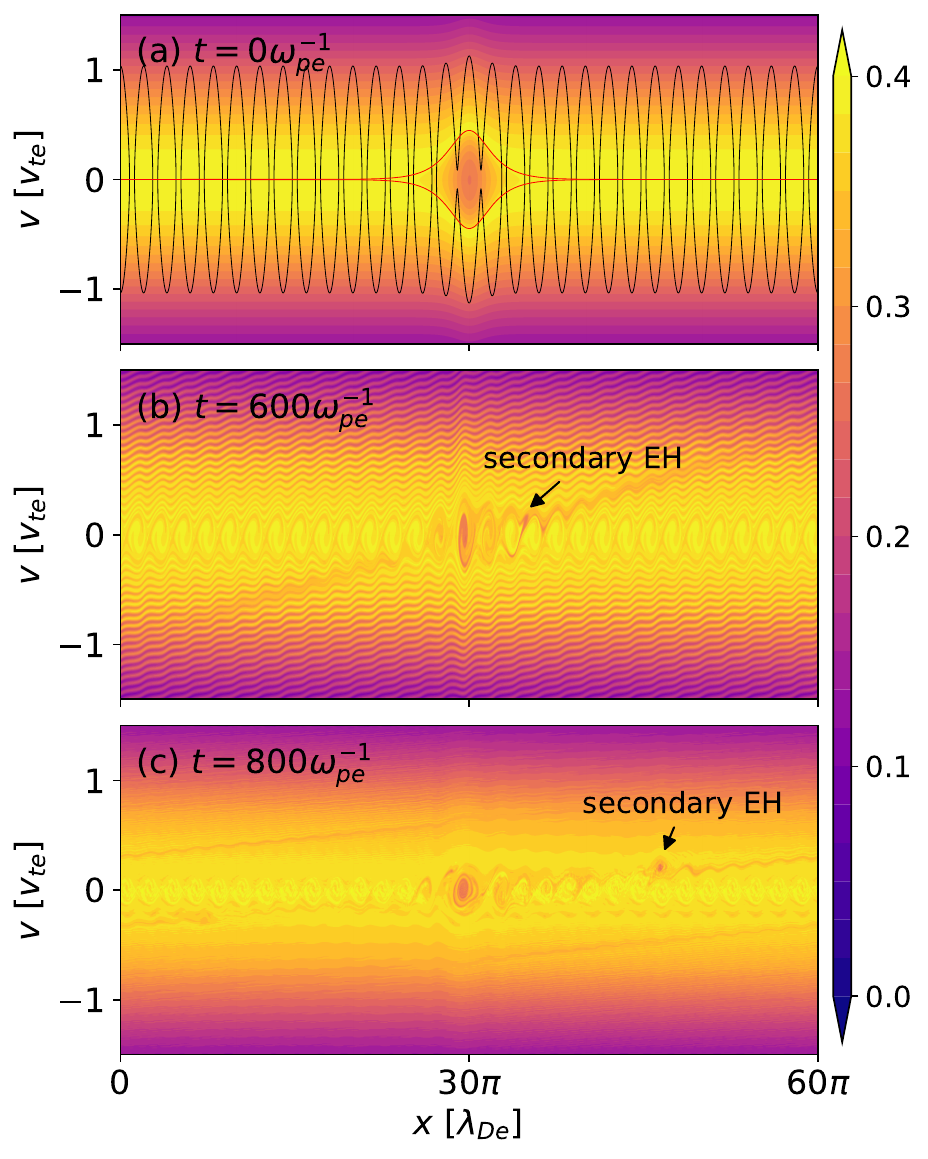}
    \caption{
        The time evolution of the electron phase space at (a) $t=0\omega_{pe}^{-1}$, (b) $t=600\omega_{pe}^{-1}$, and (c) $t=800\omega_{pe}^{-1}$ by applying the sinusoidal electric field $\mathcal{E}_a = 0.3k_BT_e/(e\lambda_{De})$ from $t_{start}=5\omega_{pe}^{-1}$ to $t_{end}=505\omega_{pe}^{-1}$.
        The red solid line denotes the separatrix between the trapped and passing electrons in the absence of the applied electric field.
        The black solid line represents the contours of constant energy after the application of the external electric field.
    }
    \label{fig:sa-sin-E03-fe-evo}
\end{figure}
In addition, similar to the case of the uniform electric field, the applied sinusoidal electric field also triggers the separation of secondary EHs from the main structure, as marked in Fig. \ref{fig:sa-sin-E03}.
The detailed evolution of the electron phase space is illustrated in Fig. \ref{fig:sa-sin-E03-fe-evo}.
In panel (a), the red solid line denotes the separatrix between the trapped and passing electrons without the external field, while the black solid lines represent the contours of constant energy immediately after the application of the sinusoidal electric field ($w = v^2/2-\phi_0-(\mathcal{E}_a/k) \cos(kx)$).
Thus, some of the originally trapped electrons can still spill out from the initial EH and form a secondary EH after the applied electric field is removed ($t > 505\omega_{pe}^{-1}$ in this simulation), as shown in Figs. \ref{fig:sa-sin-E03-fe-evo}(b) and (c).

\section{Summary}
\label{sec:summary}
This work studies the self-acceleration processes of EHs in the applied electric field by using the one-dimensional electrostatic Vlasov simulations.
At the initial stage, the EH is self-consistently constructed in the case of immobile ions in terms of Eqs. \eqref{eq:phi}-\eqref{eq:fe0-t}.
Then, the simulation starts, the ion response is turned on, and the applied electric field is adiabatically added or removed according to Eq. \eqref{eq:Ea-t}. 

The simulations are performed in three cases.
First, the benchmark simulation confirms the EH self-acceleration in the absence of the external field, and the results are plotted in Fig. \ref{fig:sa-benchmark}.
Subsequently, the different behaviors of EHs are investigated by applying the uniform and sinusoidal electric fields, respectively.
The uniform electric field imposed in the same direction as the self-acceleration can delay its occurrence and alter the final EH speed, depending on the field strength and duration, 
as shown in Figs. \ref{fig:sa-uniform-E00008}-\ref{fig:sa-uniform-E00012}.
However, the self-acceleration does not vanish in this case because the uniform electric field shifts the electron and ion velocity distributions but does not modify their local shapes.
The applied sinusoidal electric field, representing a sine wave propagating at the EH speed, can suppress the self-acceleration by locally flattening the ion velocity distribution.
Furthermore, if the applied electric field is strong enough and sustained for a sufficiently long duration, a plateau covering the velocity range of the reflected ions is formed in the ion velocity distribution. 
As a result, the ion reflections are balanced on both sides of the solitary EH,
which eventually leads to the elimination of EH self-acceleration, as illustrated in Figs. \ref{fig:sa-sin-E03}-\ref{fig:sinE-comparison}.
In addition, the applied electric field causes the separation of secondary EHs from the main EH in all simulations of this work, including both uniform and sinusoidal cases, as shown in Figs. \ref{fig:sa-uniform-E00008-fe-evo}, \ref{fig:sa-uniform-E-com-fe}, and \ref{fig:sa-sin-E03-fe-evo}.
These secondary EHs are generated by the originally trapped electrons that escape from the main EH because the additional electric field reshapes the electron phase-space trajectories.
Based on these findings, our simulations suggest that the applied electric field can significantly affect the EH self-acceleration and may play an important role in the relevant phenomena in space plasmas. 

As a final remark, the final speed of the self-accelerating EH is determined by numerous factors even in the absence of external electric fields, such as the temperature ratio $T_i/T_e$, the mass ratio $m_i/m_e$, the initial potential profile, and the initial ion distribution. \cite{Zhou2016,Zhou2018,Guillevic2025,Lotekar2020,Mandal2020} 
Therefore, a dedicated parameter scan is required to clarify the effects of these factors on the EH self-acceleration in the presence of applied electric fields, which will be explored in future work.

\begin{acknowledgments}
This work was supported by the National Natural Science Foundation of China (No.12105361).
\end{acknowledgments}

\section*{Data Availability}
The data that support the findings of this study are available from the corresponding author upon reasonable request.

\bibliography{refs}
\end{document}